# Macroscopic quantum correlation in a delayed-choice quantum eraser scheme

Byoung S. Ham
School of Electrical Engineering and Computer Science, Gwangju Institute of Science and Technology
123 Chumdangwagi-ro, Buk-gu, Gwangju 61005, South Korea
(Submitted on November 20, 2022; bham@gist.ac.kr)

**Abstract**
Quantum entanglement is known as a unique feature of quantum mechanics, which cannot be obtained from classical physics. Recently, a coherence interpretation has been conducted for the delayed-choice quantum eraser using coherent photon pairs, where phase-locked symmetric frequency detuning between paired photons plays an essential role for selective measurement-caused nonlocal correlation. Here, a macroscopic version of the nonlocal correlation is presented using orthogonally polarized optical fields in a continuous wave quantum eraser scheme in a Mach-Zehnder interferometer (MZI). The resulting polarization projection of each MZI output fields onto a rotated polarizer satisfies the violation of the cause-effect relation. Based on this macroscopic quantum eraser, the intensity product between two projected output fields satisfies the inseparable joint-parameter relation if the intensity product is selectively measured through a low pass filter to block beating signals between them.

**Introduction**
Quantum superposition is the heart of quantum mechanics whose mysterious quantum feature is in the wave-particle duality based on probability amplitudes of a single photon or particle [1,2]. In the wave-particle duality or complementarity theory of quantum mechanics, quantum measurements play an essential role in determining the undecided photon characteristics [3]. Wheeler's delayed-choice thought experiments are about wave-particle duality, where the photon's characteristics in an interferometric system are post-determined by measurements [4]. The delayed-choice quantum eraser is in regard to the violation of the cause-effect relation, where the predetermined photon's nature in an interferometric system can be retrospectively reversed by post-measurements [4,5]. Thus far, various quantum eraser experiments have been conducted using various photon natures of thermal [6], coherent [7,8], and entangled photons [9-13]. Recently, the fundamental understanding of the quantum eraser has been conducted using coherent photons to unveil the violation of the cause-effect relation [8]. Unlike common beliefs on the mysterious quantum features, the violation of the cause-effect relation is due simply to measurement-event modifications [8]. Here it should be noted that the post-measurements in the quantum eraser are not for all events of the predetermined photon characteristics but for the intentionally modified events.

   According to the EPR paradox [14], quantum entanglement implies nonlocal correlation between space-like separated paired particles, resulting in the violation of local realism [15,16]. Thus, quantum entanglement has been understood as a mysterious quantum feature that cannot be accomplished by any classical means [17]. Unlike the delayed-choice quantum eraser relating to any nature of single particles [3-10], it is commonly accepted that quantum entanglement is established between nonclassical particles such as entangled photons generated from spontaneous parametric down conversion (SPDC) processes [18-22]. In that sense, coherent photons have been automatically excluded from the potential candidates of the nonlocal quantum correlation. Recently, however, such conventional beliefs on the nonlocal correlation have been seriously challenged by a new understanding, where the wave nature of a photon can be the origin of the nonlocal quantum features [8,23]. As an example, a Franson-type nonlocal feature has been coherently analyzed for the asymmetric Mach-Zehnder interferometer using the wave nature of a photon [24]. As a result, the nonlocal quantum feature has been successfully derived using a pure coherence approach, where a coincidence detection-caused measurement modification plays an essential role, otherwise satisfies the local realism of classical physics [25].

**Results**
Figure 1(a) shows schematic of the macroscopic version of the nonlocal quantum correlation, where conventional entangled photon pairs are replaced by a typical laser light either in a continuous wave (cw) or pulse, whose paired frequencies are modulated to be opposite ($\pm \Delta f$) across the center (laser) frequency $f_0$, as shown in Fig. 1(b). This symmetrically detuned frequency pair at $f_0 \pm \Delta f$ can be generated by a pair of synchronized acousto-optic modulators (AOMs), where the detuning choices of $\pm\Delta f$ in each AOM are random for the sign but opposite between AOMs. Instead of wide bandwidth $\Delta$ as in SPDC, however, the frequency correlated pulse pair can be fixed at frequencies $f_+$ and $f_-$. Inside the MZI, each light field has random polarizations by the 45°-rotated polarizer (P). This polarizer can also be replaced by a half-wave plate (HWP) to avoid any photon loss [23]. Even though the photon characteristics inside the MZI satisfy indistinguishability resulting in interference fringes in both output fields $I_A$ and $I_B$, there is no such φ-dependent fringe in Fig. 1(a) due to orthogonal polarization bases chosen by PBS2 [26], satisfying the Fresnel-Arago law [27]. As demonstrated for a microscopic quantum version based on single photon pairs [8], the macroscopic version of the quantum eraser with optical fields in Fig. 1 can also be represented by the same Wheeler's delayed-choice thought experiments [28]. This is due to the fact that a photon never interferes with others as mentioned by Dirac [1]. In other words, there is no MZI difference between single photon and continuous wave (cw) lights, as demonstrated by cw quantum eraser [29] as well as Born rule tests [30]. Detailed analysis is presented below.

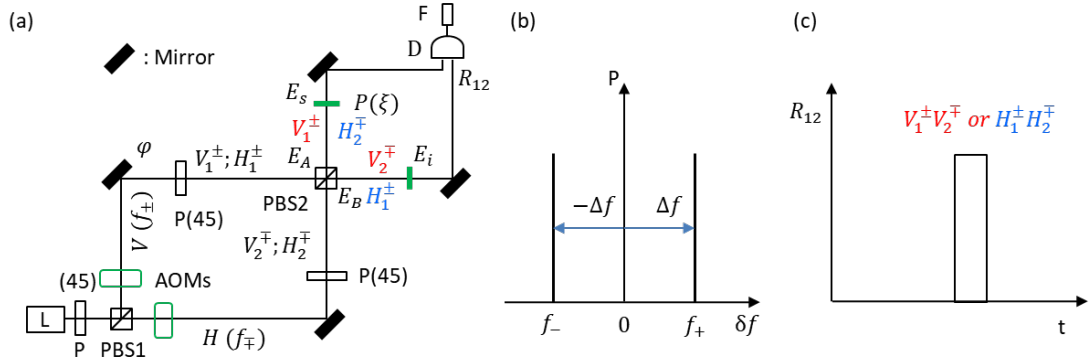

Fig. 1. Macroscopic nonlocal quantum correlation. (a) Schematic of macroscopic quantum correlation based on heterodyne detection. (b) Spectrum of a pulse pair in (a). (c) Heterodyne detection resulting intensity product in (a). BS: balanced beam splitter, AOM: acousto-optic modulator, P(45): 45° rotated polarizer, PBS: polarizing beam splitter, D: photo detector, F: low pass filter. $V_j^\pm$ ($H_j$): vertically (horizontally) polarized photon in path j, where + (-) indicates positively (negatively) detuned frequency from the laser frequency $f_0$.

The role of Ps inside the MZI in Fig. 1(a) is to satisfy indistinguishable photon characteristics in each path via random polarizations. Thus, the output measurements of $I_A$ and $I_B$ must show interference fringes if the polarizing beam splitter (PBS2) is replaced by a nonpolarizing and balanced beam splitter (BS). The added polarizer P in each output port after the PBS2 in Fig. 1(a) makes the output photon's polarizations projection onto the polarizer's rotation axis [8]. As analyzed in ref. [23], the function of the polarization projection by P is to convert the orthogonal basis into a common basis, resulting in the recovery of the indistinguishable photon characteristics [8,28]. Even though the photon loss is inevitable by P [9], the violation of the cause-effect relation is satisfied via recovering interferometric fringes. This type of cw Wheeler's delayed-choice quantum eraser has already been demonstrated using a typical laser in a macroscopic regime [28]. For the output field's joint measurements between two detectors, however, a low pass filter is added in the present macroscopic version of nonlocal correlation to exclude $\pm\Delta f$-resulting beat-frequency components. This detection-caused selective measurement is the heart of the proposed macroscopic nonlocal correlation between two independent local parameters [24]. For the practical reason of a single-photon case [8], the Ps inside the MZI can be replaced by HWPs [8].

In the MZI of Fig, 1(a), a randomly generated symmetric frequency pair by synchronized AOMs has a total of eight polarization combinations through Ps rotated by 45° via PBS2: $V_1^\pm$-$V_2^\mp$; $V_1^\pm$-$H_2^\mp$; $H_1^\pm$-$H_2^\mp$; $H_1^\pm$-$V_2^\mp$. Here, the subscript '1' and '2' indicate upper and lower paths of the MZI, respectively. The superscript '+' and '-' indicate $f_+$ and $f_-$, respectively. The MZI output pulse's amplitudes are denoted by $E_A$ and $E_B$, whose resulting polarization combinations are shown in Fig. 1(a). By PBS2, each MZI output field is represented by distinguishable polarization bases. Thus, there is no fringe in both local intensity measurements for $I_A$ and $I_B$, where $I_k = E_k E_k^*$. The analytical solutions are coherently obtained using a common method of matrix representations:

$$I_A = \frac{I_0}{4}\left(-V_1^\pm e^{i(\varphi \pm \Delta f)t} + H_2^\mp e^{\mp i\Delta f t}\right)\left(-V_1^\mp e^{-i(\varphi \pm \Delta f)t} + H_2^\pm e^{\pm i\Delta f t}\right) = \frac{I_0}{2}. \quad (1)$$

Likewise,

$$I_B = \frac{I_0}{4}\left(H_1^\pm e^{i(\varphi \pm \Delta f)t} + V_2^\mp e^{\mp i\Delta f t}\right)\left(H_1^\mp e^{-i(\varphi \pm \Delta f)t} + V_2^\pm e^{\pm i\Delta f t}\right) = \frac{I_0}{2}. \quad (2)$$

These local intensities in Eqs. (1) and (2) are of course for single shot-based as usual in coherence optics, where the P-caused intensity loss at 50 % is not included. Here, $I_0$ is the light intensity just before PBS1.

By the inserted polarizers Ps ($\xi$; $\theta$) in both MZI output ports, the output fields $E_A$ and $E_B$ are modified for polarization projections onto a common basis determined by P's rotation angles $\xi$ and $\theta$. Thus, the final amplitudes of both output fields are represented by:

$$E_s = \frac{E_0}{2}\left(-V_1^\pm \sin\xi\, e^{i(\varphi \pm \Delta f)t} + H_2^\mp \cos\xi\, e^{\mp i\Delta f t}\right), \quad (3)$$

$$E_i = \frac{iE_0}{2}\left(V_2^\mp \sin\theta\, e^{\mp i\Delta f t} + V_1^\pm \cos\theta\, e^{i(\varphi \pm \Delta f)t}\right), \quad (4)$$

where the polarization bases $V_j^\pm$ and $H_j^\pm$ are just to indicate their polarization origins. Thus, both local intensities measured in both detectors are obtained from Eqs. (3) and (4):

$$I_s = \frac{I_0}{4}\left(-V_1^\pm \sin\xi\, e^{i(\varphi \pm \Delta f)t} + H_2^\mp \cos\xi\, e^{\mp i\Delta f t}\right)\left(-V_1^\mp \sin\xi\, e^{-i(\varphi \pm \Delta f)t} + H_2^\pm \cos\xi\, e^{\pm i\Delta f t}\right),$$

$$= \frac{I_0}{2}(\sin^2\xi + \cos^2\xi - 2\sin\xi\cos\xi\cos(\varphi \pm 2\Delta f t)),$$

$$= \frac{I_0}{2}(1 - \sin 2\xi \cos(\varphi \pm 2\Delta f t)). \quad (5)$$

$$I_i = \frac{I_0}{2}\left(V_2^\mp \sin\theta\, e^{\mp i\Delta f t} + H_1^\pm \cos\theta\, e^{i(\varphi \pm \Delta f)t}\right)\frac{iE_0}{2}\left(V_2^\pm \sin\theta\, e^{\pm i\Delta f t} + H_1^\mp \cos\theta\, e^{-i(\varphi \pm \Delta f)t}\right),$$

$$= \frac{I_0}{2}(\sin^2\theta + \cos^2\theta + 2\sin\theta\cos\theta\cos(\varphi \pm 2\Delta f t)),$$

$$= \frac{I_0}{2}(1 + \sin 2\theta \cos(\varphi \pm 2\Delta f t)). \quad (6)$$

Due to the synchronized random detuning at $\pm\Delta f$ by the AOMs, the mean intensities become $\langle I_s \rangle = \langle I_i \rangle = \frac{I_0}{2}$, regardless of $\xi$ and $\theta$. If there is no randomness in $\pm\Delta f$ in each MZI path, Eqs. (5) and (6) show moving fringes as a function of time t for $\xi$, $\theta$, and $\varphi$. The half-cut intensity for both $\langle I_s \rangle$ and $\langle I_i \rangle$ with respect to $\langle I_0 \rangle$ is due to the photon loss by the last polarizer-induced projection.

Now, we solve Eqs. (3) and (4) for the joint intensity product via a low pass filter, where the joint measurements are set for exclusion of the beating signal at $2\Delta f$. This modified measurement results in a dc value, even without random polarization-basis preparation inside the MZI. This modulation frequency cut-off policy in measurements is the key mechanism of the proposed macroscopic nonlocal correlation, where all cross-colored cases in Fig. 1 need to be blocked from intensity product measurements. In the Franson-type nonlocal correlation, the common method of coincidence detection actually blocks 50% of the photons for the joint-parameter relation of nonlocality [24]. In Bell measurements by a set of polarizers, 50% photon loss is also inevitable. This selective photon measurement is the secrete to the quantum entanglement generation. We call this measurement selection quantum illusion. Without such selective measurements, no nonlocal quantum feature is observed.

The intensity product (coincidence detection) between two local detectors via the low pass filter is as follows:

$$R_{si}(\tau = 0) = (E_s E_i)(E_s E_i)^*,$$

$$= \frac{iI_0}{2}\left(-V_1^{\pm}\sin\xi e^{i(\varphi\pm\Delta f)t} + H_2^{\mp}\cos\xi e^{\mp i\Delta ft}\right)\left(V_2^{\mp}\sin\theta e^{\mp i\Delta ft} + H_1^{\pm}\cos\theta e^{i(\varphi\pm\Delta f)t}\right)(c.c.),$$

$$= \frac{I_0}{2}e^{i\varphi}\left(-V_1^{\pm}V_2^{\mp}\sin\xi\sin\theta + H_2^{\mp}H_1^{\pm}\cos\theta\cos\xi\right)(c.c.),$$

$$= \frac{I_0^2}{4}\cos^2(\xi + \theta), \tag{7}$$

where only the $V_1^{\pm}V_2^{\mp}$ and $H_1^{\pm}H_2^{\mp}$ terms are allowed. The cross-colored intensity-product components of $V_1^{\pm}H_1^{\pm}$ and $H_2^{\mp}V_2^{\mp}$ are excluded by the low pass filter. Equation (7) shows the macroscopic nonlocal fringes as functions of the joint parameters $\xi$ and $\theta$, in an inseparable intensity-product manner. Thus, the macroscopic version of the nonlocal correlation is successfully demonstrated for the proposed detection method. Surprisingly, the nonlocal correlation is nothing but measurement modification, where this measurement process is fully coherent. Because the paired coherent photons in Fig. 1(a) must have relative phase information between them, there is no difference between a single photon and a cw light for Eq. (7) (see Discussion): Neither single photons or cross-colored photon pairs contribute to the intensity product. Considering the conventional definition of classicality for coherent photons or cw light, the nonlocal correlation derived in Eq. (7) contradicts our common understanding or blind belief on quantumness. As analyzed, coherent photons should not be treated as classical particles [23-25]. In that sense, a definite phase relation between paired photons is the prerequisite for the nonlocal correlation accomplished by a measurement-event modification process.

**Discussion**
In an MZI with one input and two output ports (see Fig. 1), the interference fringes by single photons are exactly the same as cw light, as experimentally observed in 1986 [31]. This is the direct result of Born's rule [32] governed by a limited Sorkin's parameter [33]. The rule of thumb in quantum mechanics is that a photon never interferes with others [1]. Thus, Fig. 1(a) must show the same results for both single photons and cw lights in the quantum eraser [29]. Even though a BS has random distribution to the output ports, the BS-BS MZI is decisive for which port. This MZI directionality for the output photons is universal regardless of the photon characteristics whether they are thermal, coherent, or entangled. As shown in Eqs. (5) and (6), the polarizer-modified MZI outputs can be coherently manipulated for the polarization-basis control, as already demonstrated in the quantum eraser [4,8,13,29]. In Eq. (7), this coherence manipulation of independent local parameters creates the nonlocal quantum feature if the beating signal is selectively blocked off from the intensity-product (coincidence) measurements. Thus, the general concept of local randomness-based nonlocal fringe is satisfied. Because this half-selection process is common to all nonlocal correlation methods including Bell inequality violations and Franson-type nonlocal correlation, both quantum erasers and the quantum eraser-based nonlocal correlation are the so-called quantum illusion via a measurement selection process.

As mentioned above, the quantum erasers derived in Eqs. (5) and (6) should show the same results regardless of the photon nature as already demonstrated. As discussed above, the present nonlocal quantum feature is based on the quantum eraser. In microscopic coincidence detection between independently controlled MZI ports of the quantum eraser, a photon pair is a prerequisite in each measurement stage by definition of coincidence detection. In the present macroscopic regime, however, the coincidence detection method is meaningless due to the MZI directionality and automatic drop-out of unpaired single photons. The different colored-photon pairs, e.g., $V_1^\pm H_1^\pm$, are from the same MZI path violating the nonlocal condition. These cross-colored photon pairs are also automatically blocked by the low pass filter. Thus, a single photon detector is not a prerequisite condition for the present quantum measurements. The role of the low pass filter is to select only wanted photon pairs, resulting in the nonlocal correlation. As a result, the present macroscopic version of the nonlocal correlation is fully satisfied for all photon natures, as long as coherence is provided in the MZI.

**Conclusion**

A macroscopic version of the nonlocal quantum correlation was theoretically investigated in a cw delayed-choice quantum eraser scheme. Unlike the most common understanding or beliefs on the mysterious quantum feature of two-photon intensity correlation, the wave nature of quantum mechanics made it clear in terms of selective measurements. A statistical photon distribution for different-colored photon pairs in Fig. 1(a) resulted in a beat frequency. To avoid this unwanted beating signal in the cw light scheme, a low pass filter was used to select only particularly paired photons between two output ports. All single photons were also excluded from intensity product measurements by definition. By this measurement-event selection, the macroscopic nonlocal quantum correlation was established for the cw delayed-choice quantum eraser scheme, otherwise results in the classical lower bound. Thus, the present macroscopic quantum feature may open the door to macroscopic quantum information processing.

**Funding:** This research was supported by the MSIT (Ministry of Science and ICT), Korea, under the
ITRC (Information Technology Research Center) support program (IITP-2022-2021-0-01810) supervised
by the IITP (Institute for Information & Communications Technology Planning & Evaluation).

**Conflicts of Interest:** The author declares no conflict of interest.